\documentclass[%
 preprint, 
 amsmath,amssymb,
 aps,
pra,
]{revtex4-1}
\usepackage{graphicx}
\usepackage{dcolumn}
\usepackage{bm}
\usepackage{float}
\usepackage{amsfonts}
\usepackage{color}
\usepackage[english]{babel}
\usepackage[usenames,dvipsnames]{xcolor}
\usepackage{array}
\usepackage{amsmath}\usepackage{soul}
\begin{document}
\title{Multi-charged ion-water molecule collisions in a classical-trajectory time-dependent mean-field theory}
\author{Alba Jorge}
\email{albama@yorku.ca}
\author{Marko Horbatsch}
\email{marko@yorku.ca}
\author{Tom Kirchner}
\email{tomk@yorku.ca}
\affiliation{Department of Physics and Astronomy, York University, Toronto, Ontario, M3J 1P3, Canada}
\begin{abstract}
A recently proposed classical-trajectory dynamical screening model for the description of multiple ionization and capture during ion-water molecule collisions is extended to incorporate dynamical screening on both the multi-center target potential and the projectile ion. Comparison with available experimental data for He$^{2+}$ + H$_2$O collisions at intermediate energies (10-150 keV/u) and Li$^{3+}$ + H$_2$O at higher energies (100-850 keV/u) demonstrates the importance of both screening mechanisms. The question of how to deal with the repartitioning of the capture flux into allowed capture channels is addressed. The model also provides insights for data on highly-charged projectile ions (C$^{6+}$, O$^{8+}$, Si$^{13+}$) in the MeV/u range where the question of saturation effects in net ionization was raised in the literature. 
\end{abstract}
\maketitle
\section{Introduction}
The use and investigation of hadron therapy for the treatment of cancer is a promising field in current biomedical research \cite{w46,m95b,obf+07}. Even though the use of beams of heavy charged particles to attack the DNA of cancer cells has been known for many years, in the last decade more investigations have been carried out \cite{mg17}; since 2010 many operational facilities have been opened, eight of them in 2019 \cite{ptcog}.

The beams of ions interact with matter in a different way than electromagnetic waves, in the sense that the ions deposit the major part of their energy in the Bragg peak, i.e., at the end of the path they follow. This implies that the energy deposition zone can be adjusted by varying the projectile velocity and charge. Given the amount of water in the human body, the most likely event that occurs is the collision of the ion with a water molecule, which gives rise to different electronic processes such as the ejection of electrons, followed by further ionization or excitation processes \cite{sde10}. Different mechanisms after the collision can cause DNA damage, such as the creation of secondary electrons and ions, of free radicals or the heating of the medium due to target excitation \cite{tup+18}. 

The whole picture has to be considered, taking into account the effect of secondary electrons or radicals on the DNA damage \cite{ss14}. One needs an accurate understanding of the microscopic events, namely the time evolution of the involved ions leading to DNA damage \cite{bch+00}. This implies the need for atomic data such as differential and total cross sections which can then be incorporated in simulation codes \cite{cbc08}. 

Therefore, considerable attention is given to the study of collisions of different ions with biomolecules, and to water molecules in the vapor phase, since comparison of experiment and theory is feasible in this case. The main purpose is to obtain a proper atomic database \cite{dgh+13} which contains information about a number of electronic processes, such as the fragmentation of the water molecule \cite{tbv+78,wbb+95b,sph+05,lm05,cdo+07}, for different projectile ions and impact energies. 

Regarding the direct study of basic ion-molecule electronic processes, we find a variety of works dedicated to proton collisions in the literature, both experimentally \cite{tnl68,rgd+85,br86,gec+04,lbw+07,tlw+15,qmc+19} and theoretically \cite{iem+11,mkh+12b,hww+16,pte+17,tqm+18}. This is due to the fact that hadron therapy has been applied mostly with proton beams; however, it is being investigated if the use of other species which could have better physical and radiobiological properties, such as helium (alpha particles), bare carbon or oxygen ions \cite{rt17}, will provide other options. The study of collisions with such ions is more scarce but it is being developed lately for both low-charge \cite{cbc+07,lwm+16,khj+17} and high-charge projectiles \cite{nbk+13,tmf+14,bbb+16,bbm+17,obo19}. The collisions of ions with biomolecules are being investigated as well \cite{iii+13,bas18,jlh+19}.

In this paper we focus on the multi-electronic processes for ion-water molecule collisions for a variety of projectiles of interest, ranging from the proton to highly-charged ions such as Ne$^{10+}$ or Si$^{13+}$. We are interested in analyzing the importance of the many-electron aspect of the water molecule using two lines of attack. First, we look at the importance of the target and projectile potential changes due to the electron removal during the dynamics, by implementing time-dependent mean-field potentials. Second, we analyze the repartitioning of the density of removed electrons into the different multi-electronic probabilities, which is usually made through the independent particle model, i.e., trinomial analysis. The trinomial analysis can be problematic, however: six of the the water molecule's ten electrons occupy (three) weakly bound orbitals. For low-charge projectiles (protons, He$^{2+}$, Li$^{3+}$) the trinomial analysis leads to sizeable transfer probabilities for electron multiplicities that cannot be accommodated on the projectile. Therefore, we also offer an alternative analysis which does not suffer from this problem. Comparison with measurements is performed to interpret the obtained data.

We thus look at low-charge projectiles at intermediate energies, where capture and ionization compete and the effect of dynamical screening on the two centers and the repartitioning approaches are of importance. The goal is a quantitative comparison with available experimental cross sections for many channels. 

The saturation problem represents a hypothesis suggested in \cite{bbc+18} in order to explain the scaling behavior of the net ionization cross sections for high projectile charges and high impact energies. We also address this problem using the obtained data for both low and highly charged projectiles.

This paper is organized as follows; in Section \ref{teo} we explain how we have implemented the time-dependent screening in the classical trajectory method, and the new alternative analysis; Section \ref{res} is dedicated to results and analysis and the paper ends with conclusions and comments in Section \ref{conclusions}.

Atomic units ($\hbar=m_e=e=4\pi\epsilon_0=1$) are used throughout unless otherwise stated.
\section{Theoretical method\label{teo}}
This work has been implemented using the Classical Trajectory Monte Carlo (CTMC) method \cite{ap66}, where the quantum description of the electron dynamics is approximated by a classical statistical ensemble. The initial condition for this statistical model is a microcanonical distribution $\rho_M=\delta(\frac{p^2}{2}+V_{\rm{mod}}-E_{\rm{MO}})$, which is built for each of the molecular orbitals (MOs) in the water molecule. The orbital energies $E_{\rm{MO}}$ for each of the MOs are chosen as $E_{1b_1}=$ -0.5187 a.u., $E_{3a_1}=$ -0.5772 a.u., $E_{1b_2}=$ -0.7363 a.u., $E_{2a_1}=$ -1.194 a.u., $E_{1a_1}=$ -20.25 a.u. (cf. Ref.~\cite{eim+15}), and every initial ensemble contains 1$\cdot 10^5$ trajectories, which is sufficient to achieve convergence at the present level (an error of 1\% or better). While the  $1a_1$ MO plays no significant role, the $2a_1$ electrons do contribute to electron removal, albeit at a smaller scale than the weakly bound $1b_1$, $3a_1$ and $1b_2$ electrons. The limits of the classical method for the ionization process are known \cite{rb93} and are further discussed in Section III. The effective single-electron potential $V_{\rm{mod}}$ has a multi-center form \cite{rem+08,iem+11} to account for the two hydrogen and the oxygen atoms assumed to remain in the ground-state geometric arrangement. This potential takes the form:
\begin{equation} V_{\rm{mod}}=V_{\rm{O}}(r_{\rm{O}})+V_{\rm{H}}(r_{\rm{H_1}})+V_{\rm{H}}(r_{\rm{H_2}})\end{equation}
\begin{equation}
\begin{alignedat}{1}
&V_{\rm{O}}(r_{\rm{O}})=-\frac{8-N_{\rm{O}}}{r_{\rm{O}}}-\frac{N_{\rm{O}}}{r_{\rm{O}}}(1+\alpha_{\rm{O}}r_{\rm{O}})\exp(-2\alpha_{\rm{O}}r_{\rm{O}})\\
&V_{\rm{H}}(r_{\rm{H}})=-\frac{1-N_{\rm{H}}}{r_{\rm{H}}}-\frac{N_{\rm{H}}}{r_{\rm{H}}}(1+\alpha_{\rm{H}}r_{\rm{H}})\exp(-2\alpha_{\rm{H}}r_{\rm{H}})\\
\end{alignedat}
\label{pot}
\end{equation}
where $\alpha_{\rm{O}}=1.602$, $\alpha_{\rm{H}}=0.6170$. The parameters $r_{\rm{O}}$ and $r_{\rm{H}}$ represent the distances from the electron to the oxygen nucleus and the two protons, respectively. The O-H bond lengths are fixed at 1.8 a.u., and the angle between the position vectors for the protons is frozen at 105 degrees. $N_{\rm{O}}=7.185$ and $N_{\rm{H}}=(9-N_O)/2$ are the screening charge parameters for each of the centers. For each collision event we perform a rotation of the molecule with randomly distributed Euler angles to take into account all possible orientations for the target molecule. For the impact energies considered we can assume that the projectile follows a rectilinear trajectory and the rotational and vibrational degrees of freedom for the H$_2$O molecule are frozen. The collision dynamics is calculated with Hamilton's equations, and is terminated when the distance between target and projectile reaches 500 a.u.. The estimated error of the cross sections, due to the number of trajectories and the final integration time, is between 0.05 to 0.5 \% for low to medium velocities, getting to $\sim1\%$ for high-energy collisions. When the collision dynamics is finished, the single-electron probabilities for each MO $j$ are calculated as $p_j^i=n_j^i/n_{j,\rm{Tot}}$, where $i=\rm cap, ion$ stands for ionization and electron capture respectively, $n_{j,\rm{Tot}} = 10^5$ is the total number of initial trajectories, and $n_j^i$ is the number of trajectories which end the collision in each inelastic process, calculated using the following energy criterion. 
Let $E_{e-T}$ be the energy of the electron with respect to the target and $E_{e-P}$ the energy of the electron with respect to the projectile. An ionization process for each trajectory is defined as $E_{e-T}>0$ and $E_{e-P}>0$; for electron capture $E_{e-T}>0$ and $E_{e-P}<0$; and for the electron remaining in the target $E_{e-T}<0$ and $E_{e-P}>0$.
\subsection{Target and projectile dynamical response}
We study the influence of the dynamical response on the ionization and capture processes in the target and projectile potentials. In ion-atom collisions the response to electron removal and transfer has been investigated (for the target and for the projectile) \cite{khl+00,khl01,khl02,kal+05} within an independent particle framework using a time-dependent mean field. In this work we implement such ideas through the use of time-dependent target and projectile potential parameters (such as the screening charge) which depend on the net probabilities for electron removal (target potential) and electron capture (projectile potential).

Many of the theoretical details have already been presented in \cite{jhi+19} where dynamical target response was included. The time-evolution is monitored in small time-steps ($\Delta t=0.05$ a.u.) in the region where the collision happens ($t=$-10...20 a.u., where $t=0$ a.u. corresponds to the closest approach between the target and the projectile), so that the time-dependent target and projectile potentials are updated on a fine time grid. In the case of the target, the time-dependent screening has been evaluated in the same way as in \cite{jhi+19}, i.e., by making the parameters $N_O=N_O(t)$ and $N_H=N_H(t)$ dependent on the average number of removed electrons, i.e., net removal  from the target, ${\rm{P}}^{\rm{Removal}}_{\rm{Net}}(t)$. In order to do so, we rename the values $N_{\rm{O}}$ and $N_{\rm{H}}$ from Eq.~(\ref{pot}) as $N_{\rm{O}}^c$ and $N_{\rm{H}}^c$:

\begin{equation}
N_{\rm{O}}({\rm{P}}^{\rm{Removal}}_{\rm{Net}})=
\begin{cases}
	\begin{alignedat}{2}
		&N_{\rm{O}}^c\:\:\:\:\:\:\:\:\:\:&&{\rm{P}}^{\rm{Removal}}_{\rm{Net}}\leq 1\\
		&8a(1-0.1{\rm{P}}^{\rm{Removal}}_{\rm{Net}})\:\:\:\:\:\:\:\:\:\:&& 1<{\rm{P}}^{\rm{Removal}}_{\rm{Net}}\leq 10\\
	\end{alignedat}
\end{cases}
\label{potd1}
\end{equation}

\begin{equation}
N_{\rm{H}}({\rm{P}}^{\rm{Removal}}_{\rm{Net}})=
\begin{cases}
	\begin{alignedat}{2}
		&N_{\rm{H}}^c\:\:\:\:\:\:\:\:\:\:&&{\rm{P}}^{\rm{Removal}}_{\rm{Net}}\leq 1\\
		&b (1-0.1{\rm{P}}^{\rm{Removal}}_{\rm{Net}})\:\:\:\:\:\:\:\:\:\:&& 1<{\rm{P}}^{\rm{Removal}}_{\rm{Net}}\leq 10\\
	\end{alignedat}
\end{cases}
\label{potd2}
\end{equation}
where the factors $a=7.185/7.2$, $b=0.9075/0.9$ are used to make the piecewise functions continuous. 

Regarding the projectile, which enters the collision as a fully stripped ion A$^{Z_p+}$, we have implemented a model potential using the form \cite{sg74}:
\begin{equation}
V(r,t)=-\frac{1}{r}\left[\frac{N(t)}{1+H(t)({\rm{e}}^{r/d(t)}-1)}+Z_p-N(t)\right].
\label{potp}
\end{equation}
This potential was proposed to deal with neutral atoms and dressed ions, and the parameters $N$, $d$ and $H$, which determine the ionic state of the projectile, are obtained by a modified Hartree-Fock approach. In our calculations, these parameters change according to the average number of captured electrons, i.e., net capture $P_{\rm{Net}}^{\rm{Cap}}$ during the collision and therefore become functions of time during the collision. The screening charge parameter $N$ is determined according to: 
\begin{equation}
N(P_{\rm{Net}}^{\rm{Cap}})=\\
\begin{cases}
\begin{alignedat}{3}
&0				&\:\:\:\:\:\:\:\:\:\:{\rm{If}}&	\:\:\:\:\:\:\:\:\:\: P_{\rm{Net}}^{\rm{Cap}} \leq 1\\
&2(P_{\rm{Net}}^{\rm{Cap}}-1)	&\:\:\:\:\:\:\:\:\:\:{\rm{If}}&	\:\:\:\:\:\:\:\:\:\: 1 < P_{\rm{Net}}^{\rm{Cap}} \leq 2\\
&P_{\rm{Net}}^{\rm{Cap}}	&\:\:\:\:\:\:\:\:\:\:{\rm{If}}&	\:\:\:\:\:\:\:\:\:\: 2 < P_{\rm{Net}}^{\rm{Cap}} \leq Z_p\\
&Z_p				&\:\:\:\:\:\:\:\:\:\:{\rm{If}}&	\:\:\:\:\:\:\:\:\:\: P_{\rm{Net}}^{\rm{Cap}} > Z_p\:\:\:\:\:\:\:\:\:\:\:\:\:\:\: .
\end{alignedat}
\end{cases}
\label{pp}
\end{equation}

\begin{figure}
{\centerline{\includegraphics[width=0.75\linewidth]{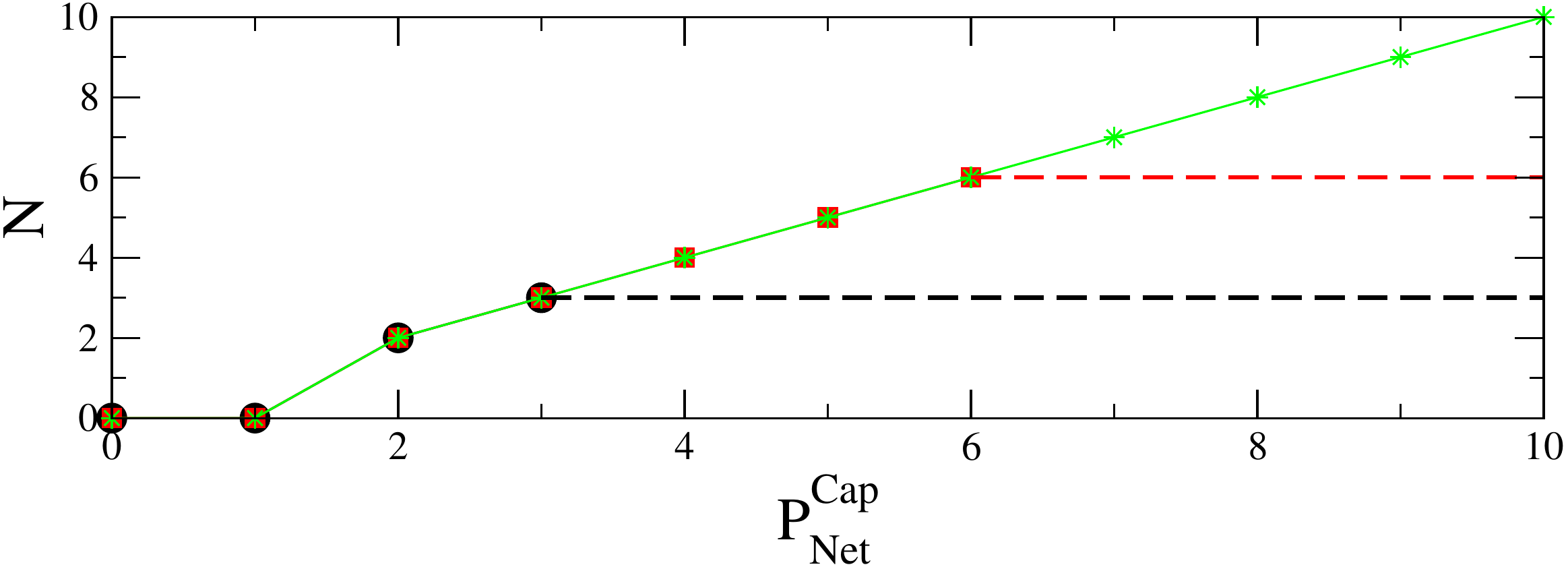}}}
\caption{Values of $N$, the screening charge of the projectile potential, for lithium (black circles), carbon (red squares) and neon (green stars) projectiles from \cite{sg74} as a function of $P_{\rm{Net}}^{\rm{Cap}}$ [Eq.~(\ref{pp})]. \label{parametros}}
\end{figure}

In Fig. \ref{parametros} the parametric dependence of $N$ is plotted versus $P_{\rm{Net}}^{\rm{Cap}}$ in accord with Eq.~(\ref{pp}). As can be seen, we only consider $N$ to go up to $N=Z_p$, which corresponds to the anion with charge -1 when the active electron is captured.
Thus, the collision starts with the Coulomb potentials for the fully-stripped ions, a potential which is maintained until $P_{\rm{Net}}^{\rm{Cap}}=2\sum_{j=1}^{5}p_j^{\rm{cap}}$ (which ranges from 0 to 10 given the five MOs with 2 electrons each) reaches the value of 1; up to this point the screening charge is 0. From this point on, the screening charge starts growing according to two linear functions which describe its dependence on $P_{\rm{Net}}^{\rm{Cap}}$ [see Eq.~(\ref{pp})]. In the first range, $1 < P_{\rm{Net}}^{\rm{Cap}} \leq 2$ the function increases from $N(P_{\rm{Net}}^{\rm{Cap}}=1)=0$ to $N(P_{\rm{Net}}^{\rm{Cap}}=2)=2$. This choice ensures consistency with the previously implemented model for the target. For $P_{\rm{Net}}^{\rm{Cap}}>2$ we model the screening charge directly by $P_{\rm{Net}}^{\rm{Cap}}$, until it reaches the value of the charge $Z_p$ of the considered ion.

In the code we calculate the net average number of removed and captured electrons at each time step using the energy criterion described
above to determine the asymptotic state, i.e., a trajectory is assigned to contribute to $P_{\rm Net}^{\rm Removal}$ if $E_{e-T}>0$ and $E_{e-P}>0$
or $E_{e-T}>0$ and $E_{e-P}<0$, and is assigned to contribute to  $P_{\rm Net}^{\rm Cap}$ if $E_{e-T}>0$ and $E_{e-P}<0$.


The values of $d_i$ and $H_i$ for the different ionic states, $i=Z_p-1,...,0$, can be found in \cite{sg74}. In the case of these parameters we have implemented piecewise functions as well, using the values for each ionic state from Table 1 in \cite{sg74} and interpolating between the integer values.
\subsection{Multinomial analysis for multiple capture evaluation}
We use the nomenclature $P_{kl}$, where the integer values $k$ and $l$ stand for the number of captured and ionized electrons, respectively. The impact-parameter-dependent probabilities for charge-state correlated channels are computed as in \cite{lwm+16},
\begin{equation}
\begin{alignedat}{1}
P_{kl}=&\sum_{k_1,...,k_5=0}^{M_1,...,M_5} \sum_{l_1,...,l_5=0}^{M_1,...,M_5}\delta_{k,\sum_i k_i}\delta_{l,\sum_i l_i}\prod_{i=1}^{5}{M_i\choose k_i+l_i}{k_i+l_i \choose k_i}\\
&(p_i^{\rm{cap}})^{k_i}(p_i^{\rm{ion}})^{l_i}(1-p_i^{\rm{cap}}-p_i^{\rm{ion}})^{M_i-k_i-l_i},
\end{alignedat}
\label{pkl}
\end{equation}
where $\delta_{k,\alpha}$ is the Kronecker delta symbol and $M_1=M_2=...=M_5=2$ refer to the number of electrons in each MO. Using this nomenclature, we define the single-capture probability $P_1^{\rm{Cap}}$ and the double-capture probability $P_2^{\rm{Cap}}$ as
\begin{equation} P_j^{\rm{Cap}}=\sum_{i=0}^{10-j}P_{ji}\:\:\:\:\:,\end{equation}
while $P_{\rm{Net}}^{\rm{Cap}}$ can also be defined as
\begin{equation} P_{\rm{Net}}^{\rm{Cap}}=\sum_{j=1}^{10}j P_{j}^{\rm{Cap}}.\end{equation}
Equation (\ref{pkl}) represents the standard independent electron model (IEM) within which cross sections for multiple processes are computed. As pointed out previously \cite{khl01,khl02}, the IEM using the trinomial analysis works well for electron removal of up to about $Z_p$ electrons, with an overestimation of high-multiplicity events. Another problem is that trinomial statistics distributes $N_t$ target electrons over three regions of space: target, projectile and continuum. However, the projectile can only accomodate $Z_p$ electrons (if the creation of a negative ion is considered an anomaly in the sense that it represents a correlated state). Following the work in \cite{mkh+12a} we construct an alternative system for computing the multi-electronic probabilities, under which the $k-$fold capture with simultaneous $l-$fold ionization processes become:
\begin{equation} P'_{kl}={Z_p\choose k}q_{l}^k(1-q_{l})^{Z_p-k}\:\:\:\:\:\:\:\:(k\leq Z_p)\label{pklprima}\end{equation}
\begin{equation} q_{l}=\frac{1}{Z_p}\sum_{k=1}^{10-l}kP_{kl}\:\:\:\:\:\:\:,\label{eqpl}\end{equation}
where $q_l$ is a single-particle capture probability while $l$ electrons are being ionized. Equation (\ref{eqpl}) is valid for most of the energy range considered in this paper. However, we have checked that for impact parameters where single-particle capture probabilities are very high (which only happens for small impact parameters at impact velocities $\lesssim 1$ a.u. with non-dynamical screening), $q_{l}$ can be slightly higher than 1. For these cases the value of $q_l$ has to be capped by unity. Equations (\ref{pklprima}) and (\ref{eqpl}) establish that
\begin{equation} \sum_{k=1}^{10-l}kP_{kl}=\sum_{k=1}^{Z_p}kP'_{kl}\:\:.\end{equation}
In the case of proton projectiles the single-capture probabiliy becomes
\begin{equation}
\begin{alignedat}{1}
P_1^{\prime\rm{Cap}}&=\sum_{i=0}^{9}P'_{1i}=P'_{10}+P'_{11}+...+P'_{19}\\
&=[P_{10}+2P_{20}+3P_{30}+...+10P_{10,0}]+[P_{11}+2P_{21}+...+9P_{91}]+...+[P_{19}]\\
&=(P_{10}+P_{11}+...+P_{19})+2(P_{20}+P_{21}+...+P_{28})+3(P_{30}+...+P_{37})+...+10(P_{10,0})\\
&=P_1^{\rm{Cap}}+2P_2^{\rm{Cap}}+3P_3^{\rm{Cap}}+...+10P_{10}^{\rm{Cap}}=P_{\rm{Net}}^{\rm{Cap}}
\:\:.
\end{alignedat}
\label{scp}
\end{equation}
For the proton case $P_2^{\prime\rm{Cap}}$ is assumed to be zero and, thus, the main problem within the IEM, which can give a rather large probability for H$^{-}$ production, and even non-zero probabilities for more highly charged anions, is removed. The transfer ionization probability $P'_{\rm{TI}}$, defined as the probablity of one electron removal accompanied by multiple ionization, is given by
\begin{equation} 
\begin{alignedat}{1}
P'_{\rm{TI}}&=\sum_{i=1}^{9}P'_{1i}=[P_{11}+2P_{21}+...+9P_{91}]+[P_{12}+2P_{22}+...+P_{82}]+...+[P_{19}]\\
&=(P_{11}+P_{12}+...+P_{19})+2(P_{21}+P_{22}+...+P_{28})+...+9(P_{91})\\
&=(P_1^{\rm{Cap}}-P_{10})+2(P_2^{\rm{Cap}}-P_{20})+...+9(P_9^{\rm{Cap}}-P_{90})\\
&=P_{\rm{Net}}^{\rm{Cap}}-i\sum_{i=1}^{10}P_{i0}\:\:.
\end{alignedat}
\label{tip}
\end{equation}
In the case of higher projectile charges the equations are less straightforward, but can be calculated using the different terms of Eq.~(\ref{pkl}). In the following section we will analyze the multiple capture results not only in terms of the importance of limiting the electron removal flux by the time-dependent screening during the dynamics, but also as a function of the multi-electronic repartitioning based on Eqs. (\ref{pklprima}) and (\ref{eqpl}). We will label the cross sections computed with the usual IEM as $\sigma$ and those computed with this alternative repartitioning approach as $\sigma'$. 
 
The total cross sections $\sigma_{i}=2\pi\int_{0}^{\infty}bP_{i}{\rm{d}}b$ follow after integration over impact parameter $b$. In practice, calculations are carried out for up to a maximum impact parameter $b_{\rm{max}}$ which is determined via the condition $P_i<10^{-5}$, where $i$ stands for ionization and capture. The maximum impact parameter changes with the impact velocity and the process considered, ranging from $b_{\rm{max}}=5$ a.u. for ionization at high velocities to $b_{\rm{max}}=8$ a.u. for capture at low velocities.
\section{Results and analysis \label{res}}
We have performed calculations for the collisions of different projectiles with the water molecule, namely H$^+$, He$^{2+}$, Li$^{3+}$, C$^{6+}$, O$^{8+}$, Ne$^{10+}$ and Si$^{13+}$ (the latter as a bare Coulomb potential). We focus on the projectiles in lower charge states to analyze the effects of time-dependent screening. We also study the repartitioning of the capture flux for these systems. The data concerning the highly-charged projectiles is used to shed light on the problem of the saturation behavior of net ionization, which has been posed in \cite{bbc+18}.

We discuss first the effect of including time-dependent potentials on both target and projectile, so that during the dynamics the change of potential parameters is taken into account. In order to compare and evaluate the importance of the screening mechanism, we include calculations with purely static potentials (no dynamical screening), and also the case where only the target response is considered. 
\begin{figure}
{\centerline{\includegraphics[width=0.75\linewidth]{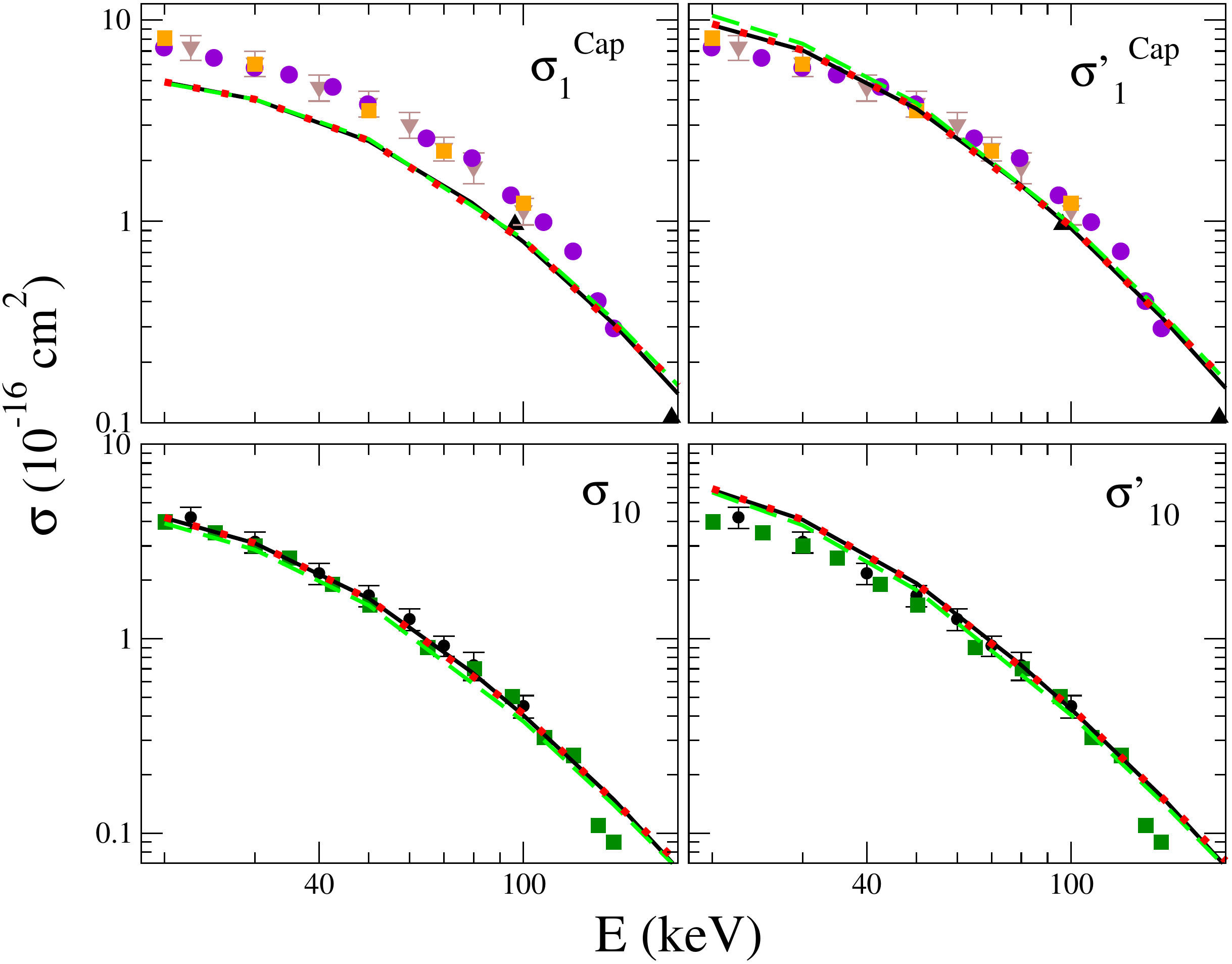}}}
\caption{In the upper panels, the single-capture cross section is shown; in the lower panels, we display the cross section for one-electron capture without ionization (i.e., pure single capture), for the collision H$^+$ + H$_{2}$O. On the left, as calculated within the trinomial analysis, on the right using the alternative repartitioning analysis. 
Present calculations: target and projectile dynamical screening (full black line), target dynamical screening (dotted red line), and purely static potential (dashed green line). Experiments, SC: black triangles \cite{tnl68}, brown inverted triangles \cite{lbw+07}; violet circles \cite{gec+04}. Net capture: orange squares \cite{rgd+85}. $\sigma_{10}$: black circles \cite{lbw+07}; green squares \cite{gec+04}. \label{figh}}
\end{figure}

We start by comparing our calculations to experimental data for proton-water collisions. In Fig. \ref{figh} we show a comparison of total single-electron capture (SC) for this system, as well as for the pure single-capture process. We have included two panels for each process so that we can compare the trinomial (standard IEM) and the alternative repartitioning approach for the calculation of multiple processes. 

As shown in Eq.~(\ref{scp}), the alternative repartitioned version for the single-capture cross section simply becomes the net cross section. Therefore, a direct comparison between the two upper panels shows that
there are non-negligible differences between the multinomial single capture and the net capture cross section. It implies that there is a substantial amount of multiple capture to the proton, which is a known problem within the binomial (or trinomial) IEM. The dynamical screening (in the two approaches) in the collision calculation does have an important effect: it is reducing $P_{\rm{Net}}^{\rm{Cap}}$ by around 30\% at 20 keV with respect to the non-dynamical screening version, but only for the lowest impact parameters. Therefore its effect on the cross section is small, but the inclusion of dynamical screening does improve the accuracy for low impact energies. Negligible differences are found between the two dynamical screening approaches for the capture process, since the reduction of the target screening charge due to the net electron removal has the effect of lowering $P_{\rm{Net}}^{\rm{Cap}}$ to values below one in almost all the cases, and therefore Eq.~(\ref{pp}) is never applied. Only for the lowest considered impact velocity and only for small impact parameters do we find a region where $P_{\rm{Net}}^{\rm{Cap}}$ is slightly higher than one when the dynamical screening is applied only on the target, but this region is sufficiently small so that no visible differences can be found between the capture cross sections computed with the two response approaches.

As can be seen in the lower panels of Fig. \ref{figh}, the effect of repartitioning the one-electron probabilities within the alternative approach for $\sigma_{10}$ leads to some disagreements with the experimental data, especially at the lowest energies. For a process involving electron removal of up to $Z_p$ electrons the IEM should work properly, and it 
is superior to the alternative analysis for the $\sigma_{10}$ channel.

\begin{figure}
{\centerline{\includegraphics[width=0.75\linewidth]{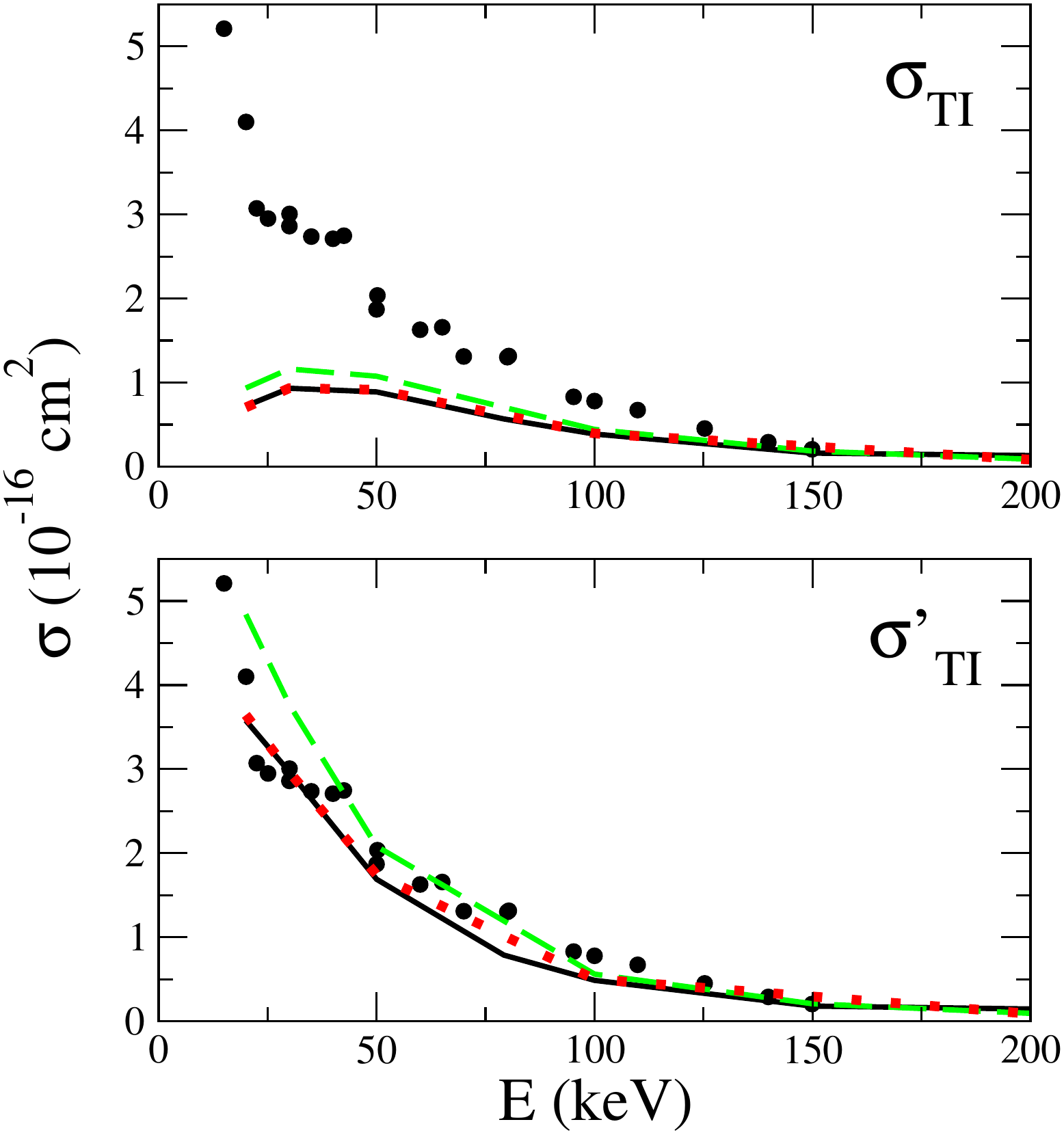}}}
\caption{Transfer ionization cross sections obtained from subtraction ($\sigma_{\rm{TI}}=\sigma_1^{\rm{Cap}}-\sigma_{10}$) using the measurements for single capture \cite{tnl68,rgd+85,lbw+07,gec+04} and for $\sigma_{10}$ \cite{gec+04,lbw+07}, for the collision H$^+$ + H$_{2}$O. Theory: in the upper panel, the transfer ionization cross section obtained within the IEM ($\sigma_{\rm{TI}}=\sum_{i=1}^9\sigma_{1i}$), in the lower panel computed as in Eq.~(\ref{tip}). The curves are denoted as: target and projectile dynamical screening (full black line), target dynamical screening (dotted red line), and purely static potential (dashed green line).\label{ti}}
\end{figure}

By using the measured data for total single capture as well as pure capture ($\sigma_{10}$), we can deduce `experimental' values for transfer ionization from the measurements in Fig. \ref{figh}, as $\sigma_{\rm{TI}}=\sigma_1^{\rm{Cap}}-\sigma_{10}$. We created a joint set of data from the different single-capture measurements, and then defined a spline function from the $\sigma_{10}$ sets of points to subtract the two quantities. In Fig. \ref{ti} we plot both the IEM and the alternative approach results (as in Eq.~(\ref{tip})) for the transfer ionization process. The improvement with the alternative repartitioning analysis in this case is especially obvious, showing again the underestimation of the single-capture cross section in the trinomial analysis, since $\sigma_1^{\rm{Cap}} = \sigma_{10}+\sigma_{\rm{TI}}$, as shown in Fig. \ref{figh}.

\begin{figure}
{\centerline{\includegraphics[width=0.75\linewidth]{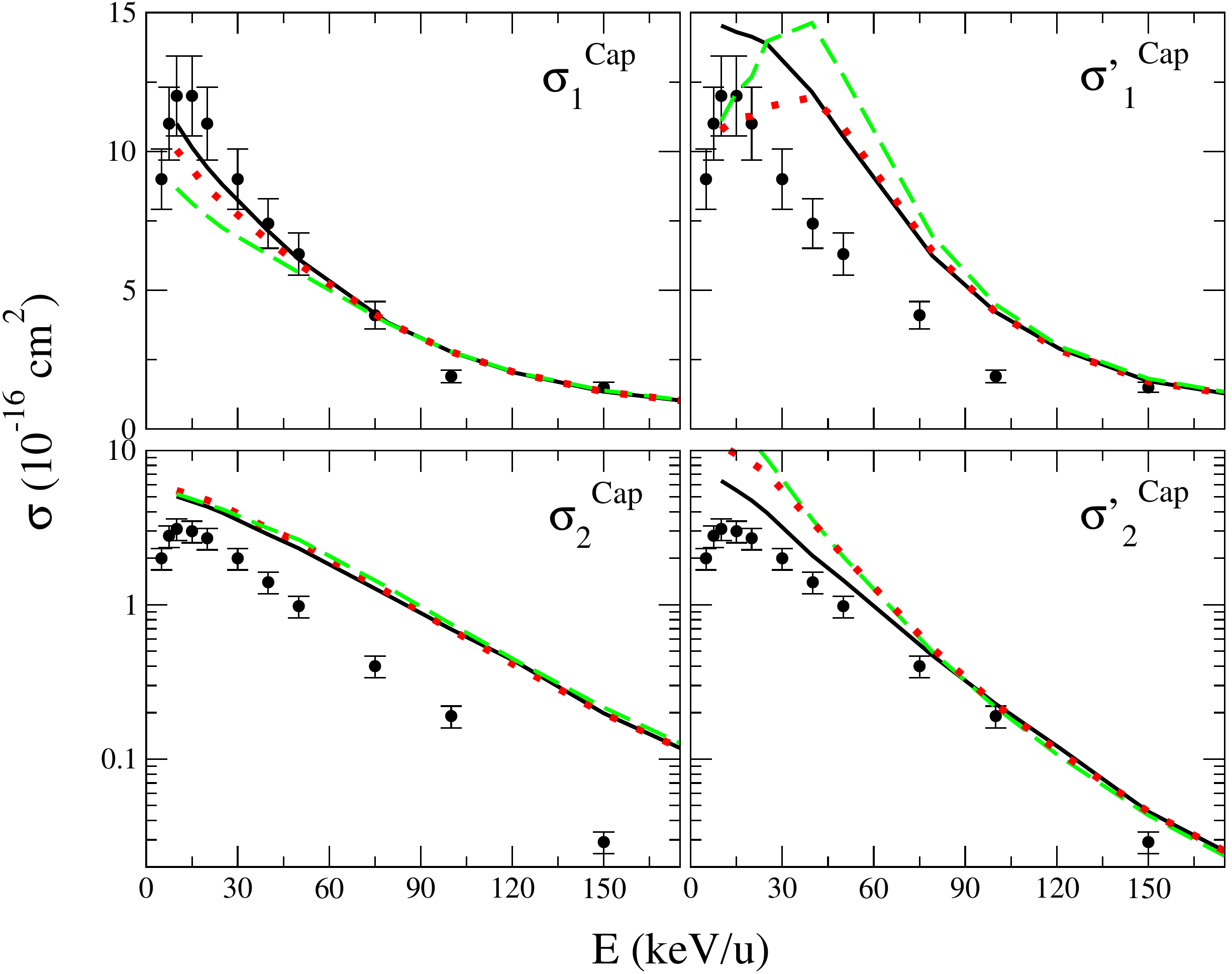}}}
\caption{Single and double capture cross section for the collisions of He$^{2+}$ with the water molecule are shown in the upper and lower panels, respectively. On the left, calculated within the IEM (trinomial analysis), on the right with the alternative repartitioning analysis. Present calculations: target and projectile dynamical screening (full black line), target dynamical screening (dotted red line), and purely static potential (dashed green line). Measurements: (black bullets with error bars) Ref. \cite{rgi85}.\label{he}}
\end{figure}

\begin{figure}
{\centerline{\includegraphics[width=0.75\linewidth]{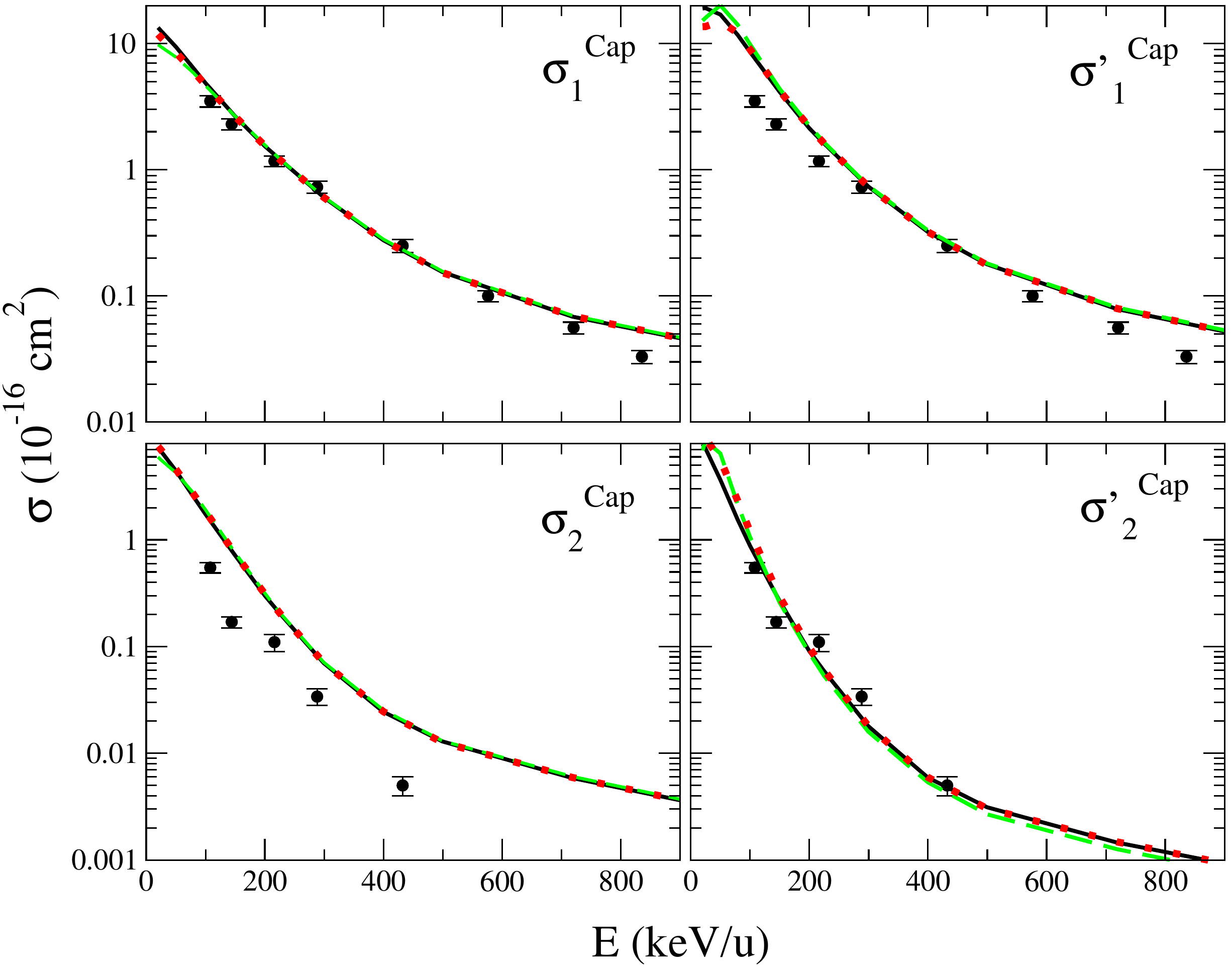}}}
\caption{Single and double capture cross section for collisions of Li$^{3+}$ with the water molecule are shown in the upper and lower panels, respectively. On the left, calculated within the IEM (trinomial analysis), on the right with the alternative repartitioning analysis. Present calculations: target and projectile dynamical screening (full black line), target dynamical screening (dotted red line), and purely static potential (dashed green line). Measurements: (black bullets with error bars) Ref. \cite{lwm+16}.\label{li}}
\end{figure}
We focus now on the comparison of results for He$^{2+}$ and Li$^{3+}$ projectiles. In Figs. \ref{he} and \ref{li} we display the single and double electron capture cross sections, including measured data from \cite{rgi85,lwm+16} and the three sets of CTMC data, computed with the multinomial and alternative analysis models. In the He$^{2+}$ system (Fig. \ref{he}), under the IEM trinomial analysis, the inclusion of dynamical screening on the projectile has an appreciable effect and it indeed improves the comparison with the experimental data in the region of low impact energy. The total effect of the inclusion of the time-dependent potentials for both target and projectile (relative to including it only in the target) in this region is the increase of the single electron capture cross section and simultaneous decrease of the double-capture cross section. When applying the alternative analysis, we find the opposite behavior in the sense that the single-capture cross section offers an inferior comparison than for double capture. It is worth noting, however, how the inclusion of dynamical screening is even more noticeable.

\begin{figure}
{\centerline{\includegraphics[width=0.75\linewidth]{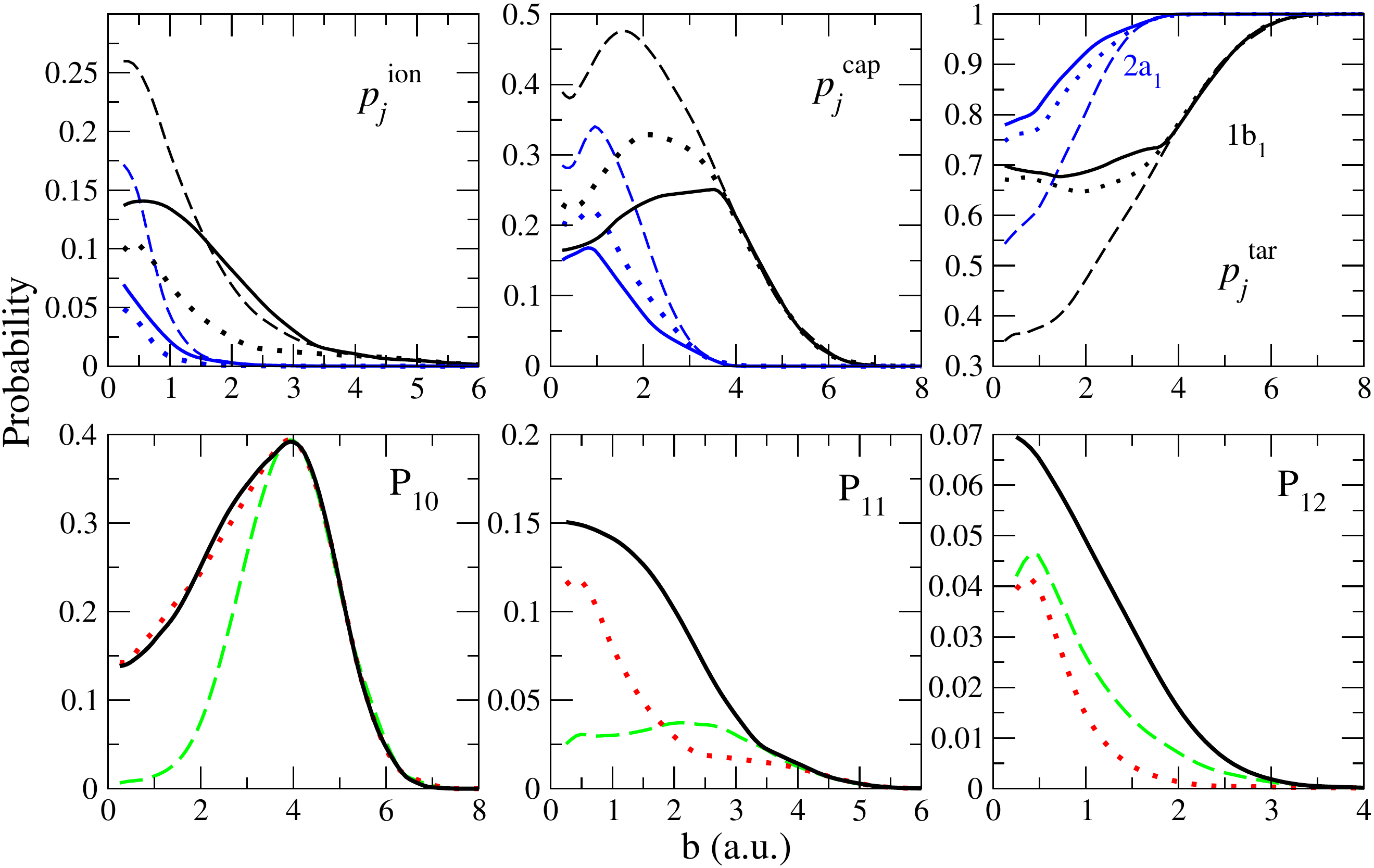}}}
\caption{In the upper panels the single-electron MO probabilities $p_j$ are shown (for 1$b_1$ in black, and 2$a_1$ in blue) from left to right for ionization, capture, and 
the complement, i.e., target probability, and in the lower panels, the calculated $P_{10}$, $P_{11}$ and $P_{12}$  for the collision He$^{2+}$ + H$_2$O at 20 keV/u. Present calculations: target and projectile dynamical screening (solid lines), target dynamical screening (dotted lines), and purely static potential (dashed lines)\label{fig-r1q3}.
}
\end{figure}

The observed increase in the single-capture cross section obtained with the dynamical screening on both centers with respect to the cross section without dynamical screening might seem counterintuitive, since the screening on the projectile implies a decrease of the single-particle capture probability. We look at an example to clarify this. The most important terms in the single-capture probability are $P_{10}$, $P_{11}$ and $P_{12}$. These processes depend strongly on the single-particle probabilities of remaining in the target, i.e., 
$p_j^{\rm tar}=1-p_j^{\rm cap}-p_j^{\rm ion}$ for $j=1,..,5$. The growth of $p_j^{\rm tar}$, as well as the decrease in $p_j^{\rm cap}$ and $p_j^{\rm ion}$ for each MO in the dynamical screening approaches imply greater values in the calculated single-capture probabilities. We show in Fig. \ref{fig-r1q3} the probabilities as a function of the impact parameter for the He$^{2+}$ projectile at the impact energy of 20 keV/u; in the upper panels the single-particle probabilities and in the lower ones, the calculated $P_{10}$, $P_{11}$ and $P_{12}$, for the three screening approaches. For the sake of clarity, we only include the 1$b_1$ and 2$a_1$ orbitals, since the results from 3$a_1$ and 1$b_2$ lie in between of those two.

For the Li$^{3+}$ case (Fig. \ref{li}), the differences between the two time-dependent screening approaches and the purely static potential are negligible for the energies considered, while using the IEM analysis for the single and double electron capture cross sections. The measurements for this projectile start at an impact energy of 100 keV/u, a region where ionization and capture do not compete anymore, and capture is much less important when compared to the region of experimental values for the He$^{2+}$ projectiles. Within the alternative analysis small differences can be found for the double-capture cross section. 

\begin{figure}
{\centerline{\includegraphics[width=0.75\linewidth]{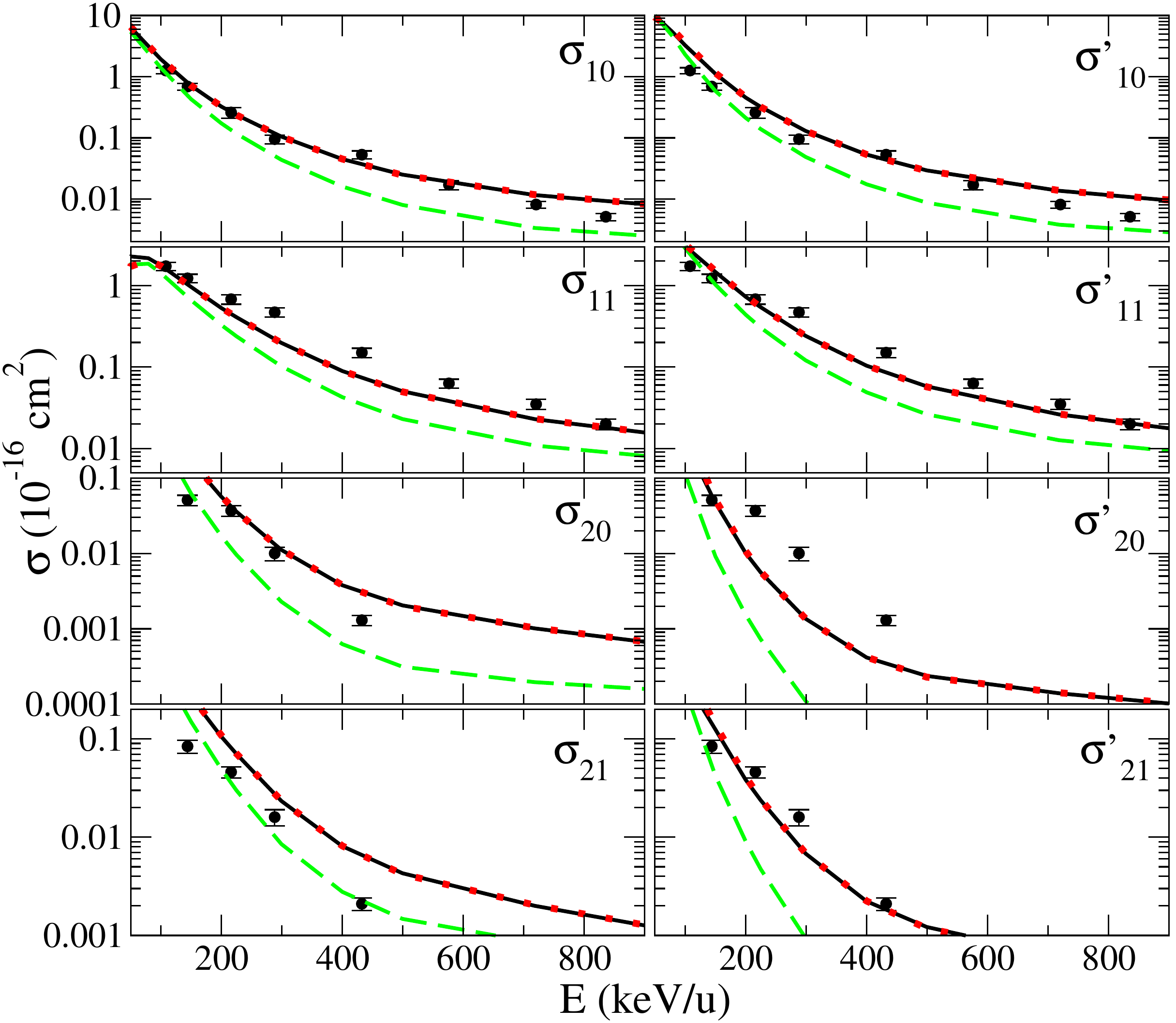}}}
\caption{Charge-state correlated cross sections for single and double capture with no ionization ($\sigma_{10}$ and $\sigma_{20}$), and with simultaneous single ionization
($\sigma_{11}$ and $\sigma_{21}$)
for  $\rm Li^{3+}-H_2O$ collisions. 
On the left, calculated within the IEM (trinomial analysis), on the right with the alternative repartitioning analysis. Present calculations: target and projectile dynamical screening (full black line), target dynamical screening (dotted red line), and purely static potential (dashed green line). Measurements: (black bullets with error bars) Ref. \cite{lwm+16}.\label{1x-2x}}
\end{figure}

Even though we find very similar results for the single-capture cross section with and without dynamical screening, the situation does not hold for the $\sigma_{1j}$ cross sections, as shown in Fig. \ref{1x-2x}. When no dynamical screening is applied the $\sigma_{10}$ and $\sigma_{11}$ are heavily underestimated especially when compared with the good agreement shown by the response data. Therefore, the no-response data predicts higher values of $\sigma_{1j}$, with $j>2$, which were not detected in the experiment \cite{lwm+16} and are not included in Fig. \ref{1x-2x}. This applies to both the IEM (trinomial) and the alternative repartitioning model.

With respect to the capture of two electrons, there is an important improvement when computed with the alternative approach, as can be seen in Fig. \ref{li}. This better comparison comes from a decrease of the capture flux for this process, which implies also a decrease of the $\sigma_{2j}$ cross sections, as shown in Fig. \ref{1x-2x}. In this case we find a better comparison with the IEM analysis for the $\sigma_{20}$ cross section and with the alternative approach for $\sigma_{21}$. It is worth noting that the most important term for the double-capture cross section is the contribution from $\sigma_{21}$ and not from $\sigma_{20}$.

\begin{figure}
{\centerline{\includegraphics[width=0.75\linewidth]{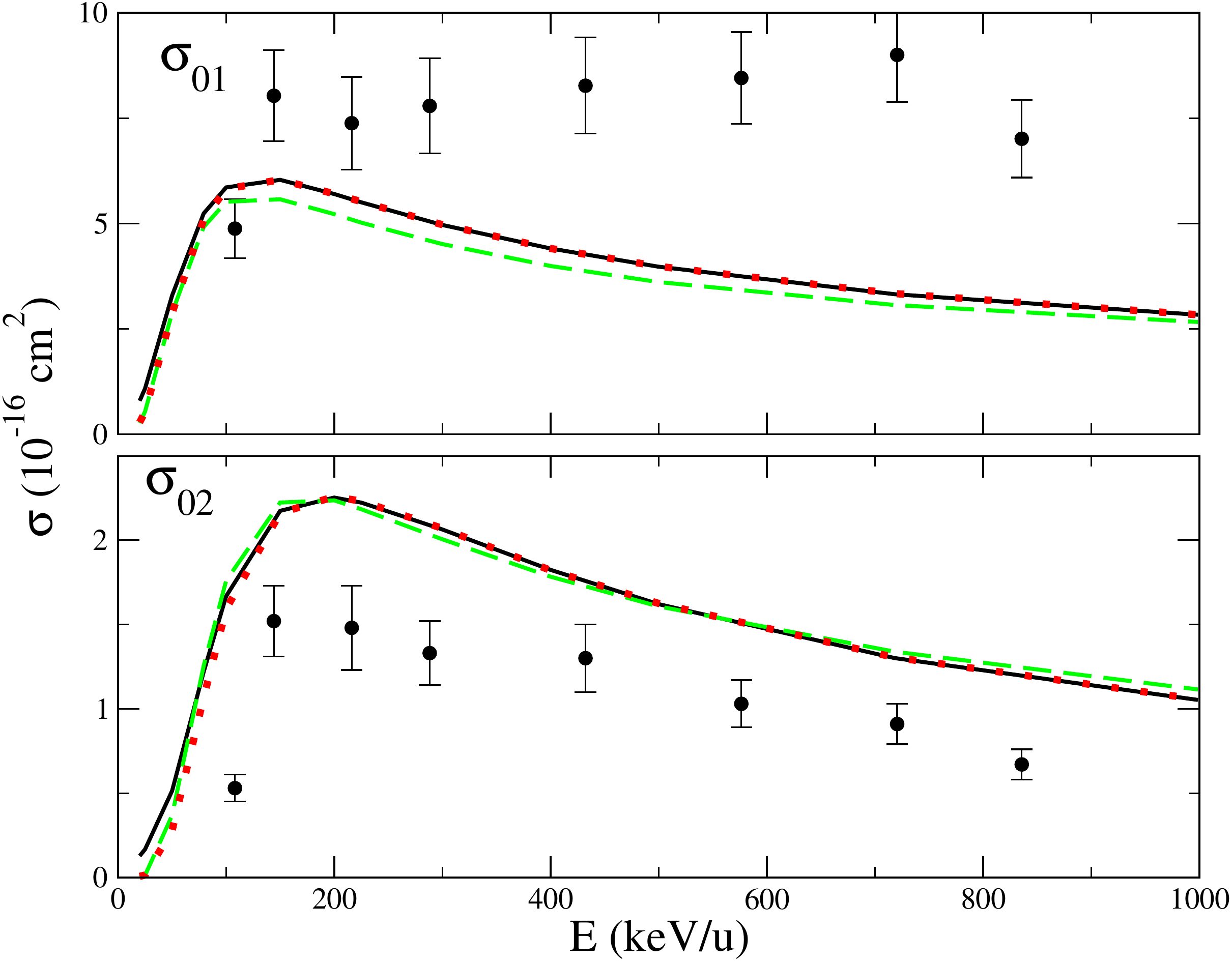}}}
\caption{Pure single and double ionization cross sections for the collision of Li$^{3+}$ with water molecules. Present calculations: target and projectile dynamical screening (full black line), target dynamical screening (dotted red line), and purely static potential (dashed green line). Measurements: (black bullets with error bars) Ref. \cite{lwm+16}.\label{0102}}
\end{figure}

We focus now on the ionization process, and therefore the alternative analysis is no longer considered and all reported results are obtained using the IEM trinomial analysis. We show the pure ionization cross sections $\sigma_{01}$ and  $\sigma_{02}$ for $\rm Li^{3+}-H_2O$ collisions in Fig. \ref{0102} to investigate differences between the three screening models in comparison with experiment ~\cite{lwm+16}. The models yield very similar results at the relatively high collision energies, and agree with the experimental data only at the factor-of-two level of accuracy, underestimating  $\sigma_{01}$ and overestimating  $\sigma_{02}$. This relatively poor performance of the CTMC models for these channels without electron capture ($\sigma_{01}$ is dominated by larger impact parameters) can be explained by the known weaknesses, such as reduced ionization probability at large impact parameters (missing quantum mechanical dipole mechanism) and possible overestimation of ionization at small impact parameters \cite{rb93}.

\begin{figure}
{\centerline{\includegraphics[width=0.7\linewidth]{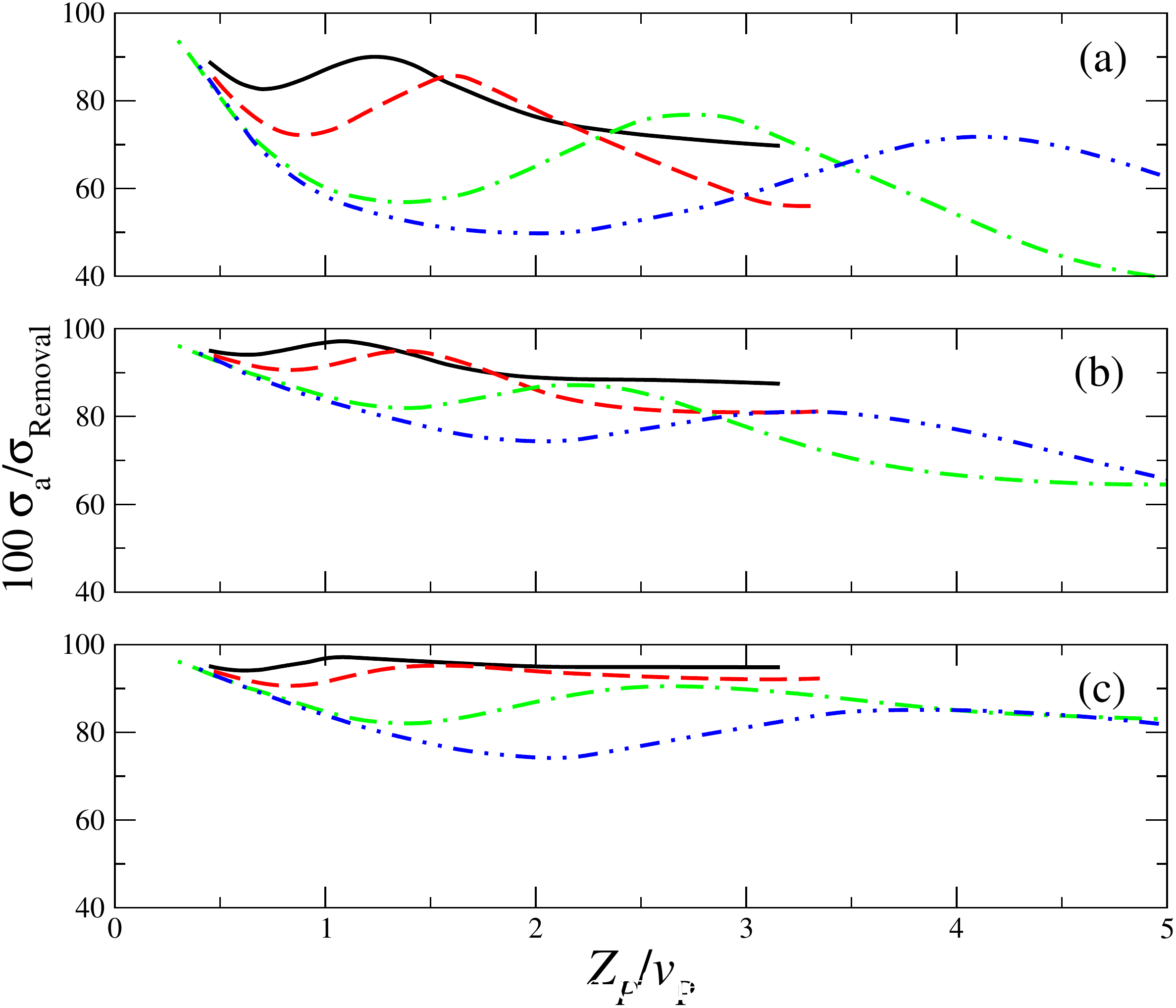}}}
\caption{Percentage of $\sum_{i=1}^3i(\sigma_i^{\rm{Cap}}+\sigma_i^{\rm{Ion}})$ with respect to $\sum_{i=1}^{10}i(\sigma_i^{\rm{Cap}}+\sigma_i^{\rm{Ion}})$ in the case of purely static potentials (a), only target response (b) and both target and projectile response (c), as a function of the Sommerfeld parameter, $Z_p/v_p$ with $v_p$ in atomic units. The systems shown are He$^{2+}$ ({\bf{-----}}), Li$^{3+}$ ({\textcolor{red}{$---$}}), C$^{6+}$ ({\textcolor{green}{$-\cdot-$}}) and Ne$^{10+}$ ({\textcolor{blue}{$-\cdot\cdot-$}}). \label{percent}}
\end{figure}

In Fig. \ref{percent} we show for the different projectiles the percentage contribution of $\sigma_a$ to the total net electron removal cross section $\sigma_{\rm{Removal}}$, where we define $\sigma_a=\sum_{i=1}^3i(\sigma_i^{\rm{Cap}}+\sigma_i^{\rm{Ion}})$ and $\sigma_{\rm{Removal}}=\sum_{i=1}^{10}i(\sigma_i^{\rm{Cap}}+\sigma_i^{\rm{Ion}})$ as a function of the Sommerfeld parameter $Z_p/v_p$ (where the projectile velocity $v_p$ is given in atomic units).  
The idea behind this presentation is to show where the high-order multi-electron processes (as predicted by theory) are of importance in order to follow up
on a discussion of the experimental data in Fig.~9 of Ref.~\cite{bbc+18}. 
A minimum in the ratio $\sigma_a/\sigma_{\rm{Removal}}$ should be interpreted as an energy zone where the high-multiplicity terms for the electron removal process are more important. The panels (a,b,c) display the difference in results for the three screening models (static, target response, target and projectile response) respectively.
The static potential model (a) shows very deep minima for projectile charges $Z_p= 3, 6, 10$ which indicates that high-multiplicity (i.e., more than three) electron removal is predicted. Panel (b) shows how this effect is reduced dramatically by target response. Additional inclusion of projectile response (c) has a small effect for $1<Z_p/v_p<3$, but does modify the results for larger values of the Sommerfeld parameter, when capture becomes the dominant target electron removal mechanism.

From Fig. \ref{percent} it can be seen that the ranges where the multiple ionization and capture processes count the most do not scale simply with $Z_p/v_p$, and that for each projectile charge this zone changes. However, we do observe a common trend for all projectiles when $Z_p/v_p$ increases from zero. All curves display a first minimum, located at different $Z_p/v_p$ values depending on the projectile charge $Z_p$. For small $Z_p$ the minimum occurs for $Z_p/v_p<1$, but for $Z_p=10$ it moves to $Z_p/v_p\sim 2$. To the right of this first minimum, we find a local maximum which is then followed by a decrease. In the zone of small $Z_p/v_p$, where all the curves tend to 100\% we approach the perturbative regime. With respect to the stationary points, the minimum is related to the zone where the ionization is the dominant process and multiple ionization processes are most important for the total net ionization cross section. For higher values of $Z_p/v_p$ than those shown in Fig. \ref{percent}, the decreasing trend of the curves is related to the same effect happening for the capture process, a region where ionization is negligible and the high multiple capture terms are more important. Quantum mechanical calculations are required in this zone. The local maximum point is related to the regime where capture and ionization processes compete.

Including time-dependent potentials which account for the ionization and capture processes during the dynamics substantially decreases the contributions from the highest multi-electronic terms to the net electron removal probabilities, and has the potential to make them more consistent with experimental observations. The role of dynamical response was tested
in ion-atom collisions, e.g., for Ne targets~\cite{khl+00,khl01}, and is deemed even more important for the water molecule with an equal number of electrons, which are, however, bound more
weakly and are more spread-out in configuration space.

\begin{figure}
{\centerline{\includegraphics[width=0.7\linewidth]{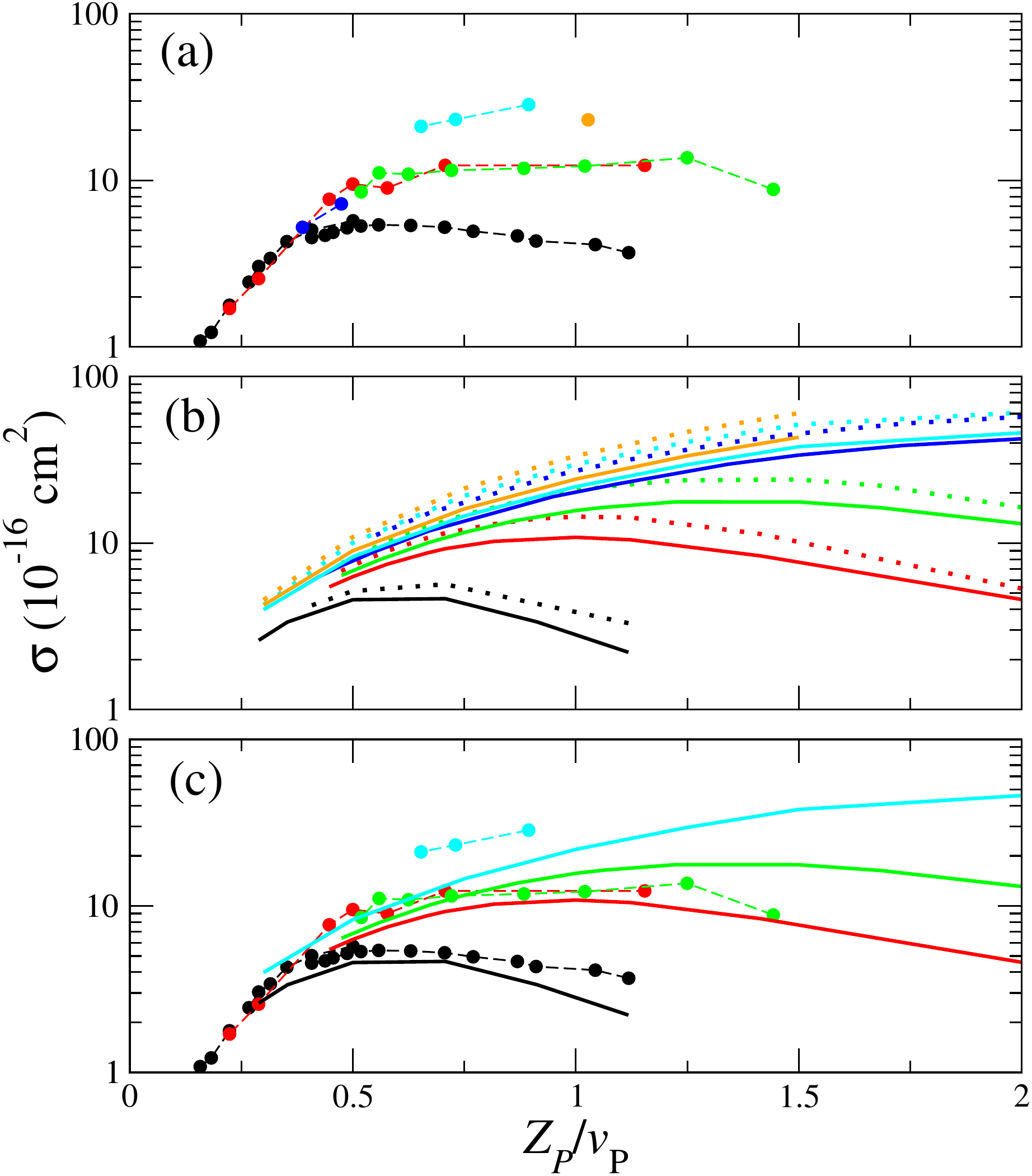}}}
\caption{In panel (a) the available measurements for the net ionization cross section as a function of the Sommerfeld parameter are shown for the following projectiles: H$^{+}$ (black, \cite{wbb+95a,gec+04,lbw+07}), He$^{2+}$ (red, \cite{twp80}), Li$^{3+}$ (green, \cite{lwm+16}), C$^{6+}$, (dark blue, \cite{dcb+09,bbm+17}), O$^{8+}$ (light blue, \cite{bbb+16,bbc+18}) and Si$^{13+}$ (orange, \cite{bbm+17}). The dashed lines connect the points to guide the eye. In panel (b), the equivalent data are given for CTMC results calculated with purely static potentials (dotted line) and both target and projectile response (full line). In panel (c), comparison of the CTMC results with response is provided with experimental data for which they are known for at least three values of $Z_p/v_p$. \label{tot}}
\end{figure}

Having identified the regions where the high-multiplicity terms for the electron removal process are important, we look now at the available experimental data for the net ionization cross sections. We plot these data in Fig. \ref{tot}, where the $x-$axis again corresponds to the Sommerfeld parameter $Z_p/v_p$. In the region $Z_p/v_p\lesssim$ 1 the saturation behavior should set in.  
The comparison of the results with and without dynamical screening shows that the net ionization cross section does not change by great amounts (typically a reduction by 30\% is observed). This seems reasonable for a global quantity which depends on the geometric distribution of the overall electron density. The comparison with experiments in panel (c) shows the need for theoretical data to assess the experimental results. As explained above the CTMC net ionization cross sections with dynamical screening are expected to approach the correct result from below since the model misses out on low-energy electrons in distant collisions. The Li$^{3+}$ + H$_2$O experimental data do not follow the expected trend as a function of $Z_p$, i.e., they are too close to the He$^{2+}$ data.

As was shown in Fig 5 of \cite{jhi+19}, in the singly differential cross sections as a function of the emission angle for the system Si$^{13+}$ ($Z_p/v_p=1.027$ a.u.), the high $q-$fold contributions are those which exhibit a more pronounced peak, while the single-ionization term shows a more decreasing shape. According to the differential measurements for the O$^{8+}$ projectile from \cite{bbb+16,bbc+18} it seems that for increasing values of $Z_p/v_p$ the ratio between the forward and the intermediate emission angles decreases, which can also be seen as a less pronounced peak with increasing $Z_p/v_p$. Therefore, experimentally for increasing values of $Z_p/v_p$ the high $q-$fold terms lose importance.

The theoretical prediction under the IEM is that the high-multiplicity terms for the ionization process become increasingly important when $Z_p/v_p$ increases, at least up to the point where for each projectile a minimum is reached in Fig. \ref{percent}. This is due to the repartitioning of the ionized flux within this model, but it still is representative of the collision itself, since the multiple electron ionization comes mainly from small impact parameters. In the case of the experimental data the figure demonstrates that for values of $Z_p/v_p\rightarrow \ 1$ single ionization becomes dominant. This could be a reason for the saturation behavior.
From our modeling point of view, as stated above, the saturation behavior is not directly related to $Z_p/v_p\rightarrow \ 1$, 
but is associated with the approach towards the first minimum as displayed in Fig. \ref{percent}(c). 
It would be of great interest to have differential measurements for a medium-high charge projectile at both the impact energy where this minimum is reached and somewhere close to it, so that this idea could be confirmed.
\section{Conclusions\label{conclusions}}
We have implemented a method to take into account dynamical response effects in both the target and projectile potentials in the CTMC description of collisions of different ions with water molecules. Calculations were carried out over a range of projectile charges $1\leq Z_p\leq 13$ covering the range of medium to high energies where the CTMC description is deemed reasonable, in order to study their influence on a qualitative level, for specific cross sections, and also to analyze their effect on the general description of multiple ionization.

Overall, the target and projectile dynamical response has been shown to yield improvements in the description of electron capture. This shows that it is important to take into account the multiple electronic processes not only through multinomial analysis, but also through the dynamics itself, for systems where a large number of electrons participate. This happens to be the case for H$_2$O, as well as for molecules of biological interest, such as the DNA and RNA nucleobases.

We have also considered two ways of analyzing the partitioning of the captured flux into the different capture channels, namely the standard IEM trinomial analysis and what we have named the alternative approach, which re-interprets the captured electron probability for a given number of ionized electrons such that only capture of up to $Z_p$ electrons is possible. This addresses a known problem with the IEM, namely the overestimation of the high-multiplicity capture channels. A downside of this analysis is a less satisfying result for the single-capture cross section compared to the IEM for $Z_p>1$. The question of how to properly distribute the captured flux remains therefore somewhat open.

In addition, this analysis has allowed us to shed light on the stated problem of the saturation behavior of the net ionization cross sections. While on the theory side we find an increase in the importance of high $q-$fold terms, when moving to high projectile charges, the experiments on Li$^{3+}$-H$_2$O do not corroborate this finding. More experimental work is clearly needed to address this question.

When comparing CTMC net cross sections as a function of Sommerfeld parameter $Z_p/v_p$ with available experimental data we observe that the latter follow the theory trend in general, but some inconsistencies remain. Thus, we are calling for additional efforts to determine normalized net ionization cross sections for ion-H$_2$O collisions.

{\bf{Acknowledgments}}

This work was supported by the York Science Fellowship program and the Natural Sciences and Engineering Research Council of Canada (NSERC). High-performance computing resources for this work were provided by Compute Canada/Calcul Canada. The authors are grateful to Hans J\"urgen L\"udde and Clara Illescas for helpful discussions.

%


\end{document}